\documentclass{actasen}

\begin{document}


\Volume{2015}{XX}


\runheading{BAO Shi-shao and SHEN Hong}%

\title{Impacts of the Nuclear Symmetry Energy on Neutron Star Crusts$^{\dag}~ \!^{\star}$}

\footnotetext{$^{\dag}$ Supported by the National Natural Science Foundation of China (Grant No. 11375089)

Received 2015--XX--XX; revised version 2015--XX--XX

$^{\star}$ A translation of {\it Acta Astron. Sin.~}
Vol. XX, No. XX, pp. XX--XX, 2015 \\
\hspace*{5mm}$^{\bigtriangleup}$ shennankai@gmail.com\\

\noindent 0275-1062/01/\$-see front matter $\copyright$ 2011 Elsevier
Science B. V. All rights reserved. 

\noindent PII: }

\enauthor{BAO Shi-shao $\quad$ SHEN Hong$^{\bigtriangleup}$ }{School of Physics, Nankai University, Tianjin 300071, China}

\abstract{Using the relativistic mean-field theory, we adopt two different methods,
namely, the coexisting phase method and the self-consistent Thomas-Fermi
approximation, to study the impacts of the nuclear symmetry energy on
properties of neutron star crusts within a wide range of densities.
It is found that the nuclear symmetry energy and its density slope play an important
role in determining the pasta phases and the crust-core transition.}


\maketitle

\section{Introduction}

It is generally believed that a neutron star mainly consists of a liquid core and crusts\rf{1}. Besides the spherical nucleus, the ones with exotic shapes, such as rod, slab, tube, and bubble, may also appear in the inner crust. That is to say, the so called pasta phases may appear there\rf{2,3}.
It has been shown in Ref.\rf{4} that the inclusion of pasta phases could
reduce the frequencies of shear modes and turn to be consistent with
observations of quasi-periodic oscillations from soft gamma repeaters.
The structure of the inner crust plays an important role in interpreting
a number of astrophysical observations, especially the giant flares in
soft gamma repeaters, neutron star oscillations, and glitches in the spin rate of pulsars\rf{5}.
In the past decades, great efforts have been devoted to study the neutron star crust properties with various methods\rf{6,7}.

The density dependence of symmetry energy is important for understanding many phenomena in both nuclear physics and astrophysics\rf{1}. The symmetry energy $E_{\rm{sym}}$ at saturation density is constrained by experiments to be $30\pm 4$ MeV, however, its density slope $L$ is still rather uncertain which may vary from about 20 to 115 MeV\rf{8}. Therefore, it is important and necessary to find a clear relation between slope $L$ and crust properties. In this paper, we systematically study the effects of the symmetry energy $E_{\rm{sym}}$ and its slope $L$ on the pasta phase properties and crust-core transition within the relativistic mean-field (RMF) theory by using two different methods, the coexisting phases (CP) method and Thomas-Fermi (TF) approximation.

\section{MODEL AND METHODS}

We adopt the Wigner-Seitz (WS) approximation to describe the inner crust matter, in which the charge neutrality and $\beta$ equilibrium conditions are satisfied. The electrons are assumed to be uniform in the WS cell because the electron screening effect may be negligible at subnuclear density\rf{9}. We use the RMF theory to describe the nucleon interaction. The Lagrangian density is given by
\begin{eqnarray}
\label{eq:LRMF}
\mathcal{L}_{\rm{RMF}} & = & \sum_{i=p,n}\bar{\psi}_i
\left\{i\gamma_{\mu}\partial^{\mu}-\left(M+g_{\sigma}\sigma\right)
-\gamma_{\mu} \left[g_{\omega}\omega^{\mu} +\frac{g_{\rho}}{2}\tau_a\rho^{a\mu}
+\frac{e}{2}\left(1+\tau_3\right)A^{\mu}\right]\right\}\psi_i  \notag \\
& & +\bar{\psi}_{e}\left[i\gamma_{\mu}\partial^{\mu} -m_{e} +e \gamma_{\mu}
A^{\mu} \right]\psi_{e}  \notag \\
&& +\frac{1}{2}\partial_{\mu}\sigma\partial^{\mu}\sigma -\frac{1}{2}%
m^2_{\sigma}\sigma^2-\frac{1}{3}g_{2}\sigma^{3} -\frac{1}{4}g_{3}\sigma^{4}
\notag \\
&& -\frac{1}{4}W_{\mu\nu}W^{\mu\nu} +\frac{1}{2}m^2_{\omega}\omega_{\mu}%
\omega^{\mu} +\frac{1}{4}c_{3}\left(\omega_{\mu}\omega^{\mu}\right)^2  \notag
\\
&& -\frac{1}{4}R^a_{\mu\nu}R^{a\mu\nu} +\frac{1}{2}m^2_{\rho}\rho^a_{\mu}%
\rho^{a\mu} +\Lambda_{\rm{v}} \left(g_{\omega}^2
\omega_{\mu}\omega^{\mu}\right)
\left(g_{\rho}^2\rho^a_{\mu}\rho^{a\mu}\right) -\frac{1}{4}%
F_{\mu\nu}F^{\mu\nu}.
\end{eqnarray}

In order to check the model and method dependence of the results obtained, we adopt the parameter sets of Bao et al.\rf{10} in the CP method, while in the TF approximation, we employ the parameter sets of Bao et al.\rf{11}. These parameter sets were generated from TM1\rf{12} and IUFSU models\rf{13}, which have the same isoscalar properties and fixed symmetry energy but have different symmetry energy slope $L$.

\section{RESULTS AND DISCUSSION}
We first present the phase diagram obtained in the CP method in Fig.1. It is found that a smaller $L$ may get more complex pasta phase structure, while only the droplet phase may appear before the crust-core transition for a larger $L$. We can see that the crust-core transition density decreases with increasing $L$.

\begin{figure}[htb]
\centering
{\includegraphics[bb=5 466 561 715,width=10cm]{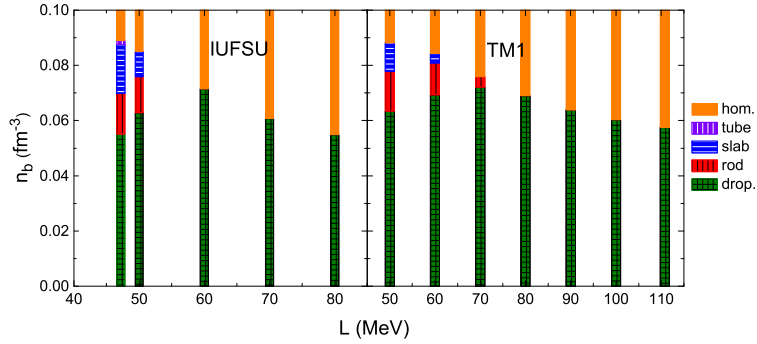}}
\caption{Phase diagram obtained in the CP method. Different colors represent corresponding pasta and homogeneous phases as indicated in the legend.}
\end{figure}

In Fig.2, the results obtained in the TF approximation are displayed. The $L$ dependence of pasta phase structure and crust-core transition is similar to the results of CP method as shown in Fig.1. But unlike the CP case, the bubble phase may appear in the TF calculation for a small $L$. It is seen that there are quantitative differences between the CP and TF methods.

\begin{figure}[htb]
\centering
{\includegraphics[bb=16 312 579 591,width=10cm]{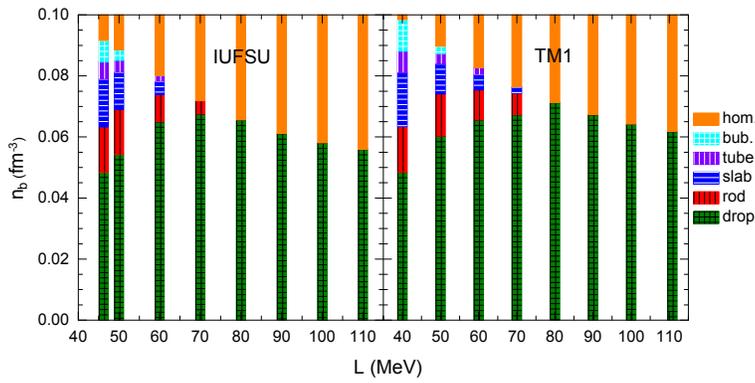}}
\caption{Phase diagram obtained in the TF approximation. Different colors represent corresponding pasta and homogeneous phases as indicated in the legend.}
\end{figure}

In this work, we consider the crusts of cold neutron stars, so the calculation is performed at zero temperature. It is interesting to study pasta phases at finite temperature in the future.

\end{document}